\documentclass[twocolumn,trackchanges]{aastex701}
\usepackage{amsmath}
\usepackage{CJK}
\usepackage{graphicx}

\definecolor{HHcolor1}{rgb}{0.93,0.57,0.13}
\definecolor{HHcolor2}{rgb}{0.5,0.1,0.5}
\definecolor{HHcolor3}{rgb}{0.0,0.0,1.0}
\definecolor{HHcolor4}{rgb}{1.0,0.0,0.0}

\newcommand{\BlueTxt}[1]{{\color{HHcolor3}#1}}
\newcommand{\RedTxt}[1]{{\color{HHcolor4}#1}}

\begin{document}
\begin{CJK*}{UTF8}{gbsn}
\title{Back to Normal Again: Possible Destinies of JWST overmassive SMBHs and ``Little Red Dots'' \\ in the View of Shin-Uchuu Simulation}

\author[orcid=0000-0003-3143-3995,sname='Hu']{Haojie Hu (胡豪杰)}
\altaffiliation{JSPS Fellow}
\affiliation{Center for Computational Science, University of Tsukuba, 1-1-1 Tennodai, 
Tsukuba, Ibaraki 305-8577, Japan}
\email[show]{huhaojie@ccs.tsukuba.ac.jp}
\correspondingauthor{Haojie Hu (胡豪杰)}

\author[0009-0006-6763-4245]{Hiroto Yanagisawa}
\affiliation{Institute for Cosmic Ray Research, The University of Tokyo, 5-1-5 Kashiwanoha, Kashiwa, Chiba 277-8582, Japan}
\affiliation{Department of Physics, Graduate School of Science, The University of Tokyo, 7-3-1 Hongo, Bunkyo, Tokyo 113-0033, Japan}
\email{yana@icrr.u-tokyo.ac.jp}

\author[0000-0003-4321-0975]{Moka Nishigaki}
\affiliation{National Astronomical Observatory of Japan, 2-21-1 Osawa, Mitaka, Tokyo, 181-8588, Japan}
\affiliation{Department of Astronomical Science, The Graduate University for Advanced Studies, SOKENDAI, 2-21-1 Osawa, Mitaka, Tokyo, 181-8588, Japan}
\email{moka.nishigaki@grad.nao.ac.jp}

\author[orcid=0009-0004-4332-9225]{Tomokazu Kiyota}
\affiliation{Department of Astronomical Science, The Graduate University for Advanced Studies, SOKENDAI, 2-21-1 Osawa, Mitaka, Tokyo, 181-8588, Japan}
\affiliation{National Astronomical Observatory of Japan, 2-21-1 Osawa, Mitaka, Tokyo, 181-8588, Japan}
\email{tomokazu.kiyota@grad.nao.ac.jp}

\author[orcid=0000-0002-5316-9171]{Tomoaki Ishiyama}
\affiliation{Digital Transformation Enhancement Council, Chiba University, 1-33, Yayoi-cho, Inage-ku, Chiba, 263-8522, Japan}
\email{ishiyama@chiba-u.jp}

\author[orcid=0000-0002-2309-3639,sname='Ohsuga']{Ken Ohsuga}
\affiliation{Center for Computational Science, University of Tsukuba, 1-1-1 Tennodai, 
Tsukuba, Ibaraki 305-8577, Japan}
\email{ohsuga@ccs.tsukuba.ac.jp}
 
\begin{abstract}
The James Webb Space Telescope (JWST) has enabled the discovery of hundreds of supermassive black holes (SMBHs) at redshifts $z\gtrsim 4-7$. A non-negligible fraction of these SMBHs are hosted in galaxies with BH-to-galaxy mass ratios ($M_{\rm BH}/M_\star$) being excessively larger than that for local SMBHs by $\sim 1-2$ dex. The origin of these ``overmassive'' BHs remains elusive, demanding either a heavy seed formation scenario or rapid growth of seed BHs. Their deviation from local scaling relations challenges our understanding of how SMBHs and their host galaxies coevolve across cosmic time. In this paper, we apply phenomenological modelings for BHs and galaxies to dark matter halo merger histories from N-body simulations to investigate the subsequent evolution of JWST-discovered ``overmassive'' SMBHs. We find that early evolution of ``overmassive'' SMBHs is dominated by stunted accretion leading to gradual decreases in $M_{\rm BH}/M_\star$ ratios. In contrast, less massive SMBHs experience super-Eddington accretion during their early evolution, resulting in a slow increase of mass ratios toward $M_{\rm BH}/M_\star \sim 0.01$. Convergence occurs at $M_{\rm BH}\sim 10^8~M_\odot$ with $M_{\rm BH}/M_\star \sim 0.01$. At lower redshift, nearly all SMBHs evolve onto local relations, as expected given that our models adopt empirical relations derived from low-redshift observations. This suggests that the global feedback mechanisms regulating the coevolution of $M_{\rm BH}/M_\star$ ratios are implicitly encoded in local relations in terms of star-formation rate distribution, black hole accretion rate distribution and their active (quiescent) fractions.
\end{abstract}


\section{Introduction}
\label{sec:intro}

Supermassive black holes (SMBHs) is ubiquitously found to exist in massive galaxies both in local Universe and at cosmic dawn \citep[][]{Ho+2008,Greene+2020,Fan+2023}. The presence of SMBHs with masses of order $10^9$ solar mass ($M_{\rm BH}\geq 10^9~M_\odot$) at $z\sim 7$ has been one of the momentous discoveries in past decades \citep[e.g.,][]{Mortlock+2011,Wu+2015,Banados+2018}, which challenges and shapes our understandings on the early Universe. However, the advent of the James Webb Space Telescope (JWST) is revolutionizing the scenario via its superb sensitivities at infrared bands. It has enabled the discovery of SMBHs to fainter luminosity end and at deeper Universe. Up to now, hundreds of SMBHs with masses $M_{\rm BH}\sim 10^6-10^9~M_\odot$ at redshift $z\sim 4-11$ haven been identified \citep[][]{Onoue+2023,Harikane+2023,Maiolino+2024,Kocevski+2025}. A significant fraction of these newly discovered SMBHs are hosted in galaxies (if any) with BH-to-galaxy mass ratios (hereafter MM ratios) reaching $M_{\rm BH}/M_\star\sim 0.01-0.1$ \citep[][]{Kocevski+2025}, values significantly higher than the empirical value observed in the present day Universe by more than an order of magnitude \citep[][]{Chen+2025}. They are dubbed as ``overmassive'' SMBHs due to their high mass ratios. Intriguingly, many of these ``overmassive'' SMBHs are found in ``Little Red Dots'' \citep[LRDs,][]{Matthee+2024}, a population of high-$z$ objects named after their unique compactness in size and redness in spectral energy distribution (SED) colors \citep[known as ``V-shaped'' SED,][]{Matthee+2024,Yue+2024,Greene+2024,Maiolino+2024,Maiolino+2024b,Stone+2024,Labbe+2025,Stone+2025,Jones+2025,Kocevski+2025,Napolitano+2025}. The nature of LRDs remains elusive, but are preferentially believed to be 
active galactic nuclei \citep[AGNs, e.g.,][]{Kido+2025}. The bulk populations of LRDs are found to raise and disappear fast at redshifts $4\lesssim z \lesssim 7$ \citep[][]{Inayoshi+2025,Kocevski+2025}, further blurring their natures. These findings challenge and re-shape our understanding of the evolution of SMBHs and galaxy formation in the cosmological context.

To coordinate the presence of massive SMBHs at cosmic dawn, two main scenarios have been proposed \citep[e.g.,][and references therein]{Inayoshi+2020,Volonteri+2021}: (1) rapid growth of light seed BHs via super-Eddington accretion with a high duty cycle
from $z \gtrsim 20$ \citep{Madau+2001, Haiman+2001,Volonteri+2003}, and (2) moderate (sub-Eddington) growth of massive seed BHs originating from either direct collapse of massive primordial gas clouds (so called DCBHs) or runaway stellar collisions in dense environments \citep[e.g.,][]{Bromm+2003,Regan+2009a, Regan+2009b,Li+2023a}. Without additional constraints, both scenarios are found to be probable. However, the newly discovered high mass ratios JWST SMBHs $M_{\rm BH}/M_\star\gtrsim 0.01-0.1$ provide severe constraints on the formation and evolution of high-$z$ SMBHs, favoring the massive seed BH origins \citep[][]{Maiolino+2024b,Bogdan+2024,Natarajan+2024}. Meanwhile, it has also been pointed out that intermittent super-Eddington accretion could also drive even light seed BHs to outgrow their hosts if sustained gas inflows are available \citep[][]{Inayoshi+2022a,Hu+2022b,Scoggins+2024,Hu+2025a}. The degeneracy comes from the treatment of feedback, being either successful or failed in boosting the rapid evolution of seed BHs. Regardless of their formation, high-$z$ SMBHs must evolve to $z=0$ in a way to manifest the observed scaling relations of $M_{\rm BH}-M_\star$. It raises the question that if these ``overmassive'' SMBHs will evolve to the local relations or not and how feedback regulates their evolution in $M_{\rm BH}-M_\star$ relation as a function of redshifts. It has been partially addressed in some cosmological simulations and theoretical modelings. For instance, \citet{Stone+2025} pointed out that the object ULAS J1120+0641 may never reach the local relations based on its gas reservoir estimations, while cosmological simulations always predict an ``undermassive'' phase evolution \citep[][]{Zhu+2022,Valentini+2021,Bhowmick+2025,Seeyave+2025,Schaye+2025,Ma+2025}. Although numerical simulations are powerful tools, the large amount of free parameters and expensive computational costs impede a clear and complete interpretation of the main drivers for the evolution of BHs and galaxy formation.

The redshift evolution of MM ratio has been discussed extensively in observation, while no consensus has been reached on whether the $M_{\rm BH}-M_\star$ relations are redshift dependent due to differences in sampling and analyzing approaches \citep[][]{Decarli+2010,Bennert+2011,Matsuoka+2018,Li+2022,Li+2025}. In light of JWST, it is intriguing to investigate the redshift evolution of the mass ratio for ``overmassive'' SMBHs in semi-analytical modeling and shed lights on feedback effects. 
In this paper, we investigate the possible evolution of JWST ``overmassive'' SMBHs and LRDs based on dark matter (DM) halo merger trees from N-body simulations and phenomenological modeling for BH and galaxy evolution. In Section~\ref{sec:model}, we lay out the modeling for DM merger histories, galaxy formation models, BH growth model and our modeling process. We present our modeling results in Section~\ref{sec:LRDs} and discuss implications and feedback models in Section~\ref{sec:Discussion}. In Section~\ref{sec:conclusions}, we summarize our conclusions. Throughout the paper a flat $\Lambda$ cold dark matter ($\Lambda$CDM) universe is adopted with cosmological parameters: $h = 0.6732$, $\Omega_{\rm m}= 0.3158$, and $\sigma_{8} = 0.8120$\citep[][]{Planck+2020}.

\section{The Shin-Uchuu Simulation and Phenomenological Modeling} \label{sec:model}

In this study, we construct a semi-analytical framework to model the possible subsequent evolution for high-$z$ SMBHs and LRDs, based on different dark matter merger histories extracted from Shin-Uchuu N-body simulation \citep[][]{Ishiyama+2021,Aung+2023,Oogi+2023}. We adopt the BH growth model from \citet[][]{Hu+2022b,Hu+2025a}, where super-Eddington accretion onto nuclear BHs are considered. To reconcile the galaxy formation constraints, galaxy evolution is modeled adopting a phenomenological modeling where star formation rate (SFR) and galaxy quiescent fraction are drawn from observations in the local Universe and at moderate low redshifts \citep[][]{Speagle+2014,Leja+2022}. Combining all the three components, we systematically model the subsequent evolution of high-$z$ ($z\geq 6$) JWST SMBHs and LRDs\footnote{For comparison, some ``low''-mass-ratio objects are also included, e.g., GN-z11 and CEERS-20496.}.

\subsection{Dark Matter Halo Merger Trees}\label{sub:DMHalo}

Dark matter halos are undoubtedly the most fundamental evolutionary arenas where galaxies and BHs build their masses. The mergers between dark matter halos act to reshape these energetic arenas and lay significant impact on the evolution between galaxies and SMBHs. Despite violent baryonic effects, dark matter evolution is fairly modeled in N-body simulations, e.g., Millennium and their descendants \citep[][]{Springel+2005}. In this paper, we adopt dark matter halo evolutions depicted in one of the largest high-resolution cosmological N-body simulations, i.e., Shin-Uchuu simulation \citep[][]{Ishiyama+2021,Aung+2023,Oogi+2023}. The highest resolution simulation of the suite, named Shin-Uchuu, consists of $262$ billion ($6400^3$) particles in a box of side-length $140 ~h^{-1}~{\rm Mpc}$, with particle mass of $8.97 \times 10^5 ~h^{-1} M_\odot$. The large simulation box and high-resolution natures of Shin-Uchuu simulation suite have made it possible to trace the halo evolution upto $z\sim 14$ for a wide range of masses, making it the ideal target to investigate the cosmic evolution for various high-$z$ SMBHs and LRDs. 
We caution that the cosmological parameters adopted in Shin-Uchuu simulations \citep[][]{Planck+2015} are slightly different from that adopted in this paper. The differences are fairly small, and no conversion between the two data sets is conducted.

\subsection{Galaxy Formation Model} \label{sec:galaxy}

Apart from dark matter halo modeling, galaxy formation is tightly connected to various interactions between different baryonic components. These baryonic processes have impeded a direct and comprehensive modeling of galaxy formation and evolution. Despite potential complexities, various attempts have been made to outline the galaxy formation. Among all attempts, analytical models and high-resolution cosmological simulations have been successful in addressing some aspects of galaxy formation. However, they either suffer from vast parameterizations or huge computational expenses. In this paper, we circumvent these caveats and adopt a phenomenological modeling for galaxy evolution to capture the mass growth for both star-forming galaxies and quiescent galaxies, aligning with observational constraints.

\begin{figure}[t!]
    \centering
    \includegraphics[width=0.85\linewidth]{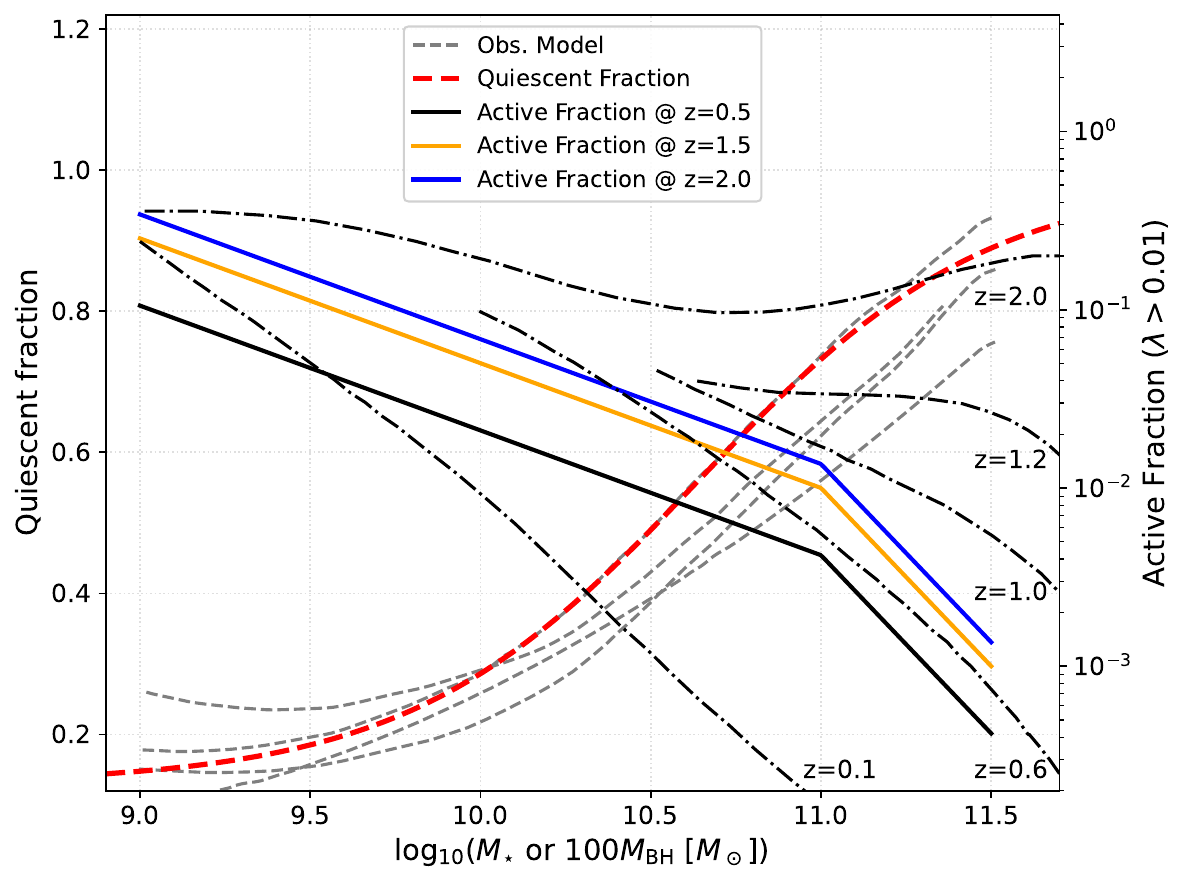}
    \caption{This figure illustrates functional forms for quiescent fractions of galaxies (dashed curves) and active fraction of BH activities (with Eddington ratio $\lambda > 0.01$, solid curves). For quiescent fraction, observational constraints inferred from different methods are overlaid as gray dashed curves for comparisons \citep[][]{Leja+2022}. For BH active fraction, models from observational indications at different redshift bins are overlaid in dash-dotted curves \citep[][]{Schulze+2015}. 
    For simplicity, a two-slope functional form with moderate redshift evolution for active fractions is adopted (solid curves). The absolute values of active fraction at high mass end is small, the exact function forms won't modify our results significantly.}
    \label{fig:QFAF}
\end{figure}

In our modeling, the galaxy mass evolution is the main focus of our interests. The galaxy gains its mass via star-formation processes and loses mass due to stellar evolution. To keep our modeling concise and statistically reasonable, SFR is modeled in a purely empirical manner and the mass loss is fairly captured as did in analytical modeling \citep[e.g., ][]{Moster+2018}. The straight-forward modeling of mass evolution allows for essential investigation of feedback effects between star-formation and BH activities (see discussions in Section~\ref{sec:Discussion}). The galaxy evolution is characterized with stellar mass evolution. Thus, baryonic processes with no contributions to star-formation and stellar mass loss are neglected in this phenomenological modeling. The star formation rate is drawn from the empirical main-sequence relation for star-forming galaxies \citep[][]{Speagle+2014}:
\begin{equation}
\log {\rm SFR}(M_\star, ~t) = (\alpha - \beta\cdot t) \log M_\star - (\alpha_0 -\beta_0 \cdot t),
\label{eq:SFR}
\end{equation}
where $M_\star$ is stellar mass of the galaxy and $t$ the cosmic age, and with coefficients: $\alpha = 0.84^{+0.02}_{-0.02}$, $\beta=0.026^{+0.003}_{-0.003}$, $\alpha_0=6.51^{+0.24}_{-0.24}$ and $\beta_0=0.11^{+0.03}_{-0.03}$. For any given stellar mass $M_\star$ at redshift $z(t)$, a SFR will be generated from the log-normal distribution $N(\log {\rm SFR}(M_\star,~t),~\sigma)$ with the $\sigma=0.3$ the observational scatter. If the galaxy is categorized as a star-forming galaxy (e.g., ${\rm sSFR} < 10^{-10}~{\rm yr}^{-1}$ etc.), it forms stars with ${\rm SFR}(M_\star,~t)$, while for a passive galaxy, the mass evolution evolves with $10^{-1.5}\cdot {\rm SFR}(M_\star,~t)$.
Since details of a galaxy are neglected, the activeness of a galaxy is determined by the ``quiescent fraction''. It has been reported that the fraction of quiescent galaxies shows nearly monotonic increases with increasing stellar mass \citep[e.g., ][]{Leja+2022}, while the redshift evolution is not significant and contributing a $20\%$-level variance. Thus, we adopt a $\tanh$ fitting to the quiescent fraction outlined in \citet{Leja+2022}:
\begin{equation}
    f_{\rm q} = f_{\rm q,~0} + k_{\rm q}\cdot \tanh\left( \frac{\log M_\star-\log M_{\rm q}}{\delta \log M_0}\right),
    \label{eq:Fquiescent}
\end{equation}
with coefficients: $f_{\rm q,~0} = 0.56$, $k_{\rm q}=0.43$, $\log M_{\rm q}=10.64$ and $\delta \log M_0=0.85$.
This treatment of SFR for star-forming and quiescent galaxies reconciles the bimodal distribution of SFR for local galaxies (see Figure~\ref{fig:QFAF} for comparison between this model and observational indications). This modeling also allow us to successfully build the evolution for host galaxies that meets the constraints from stellar-to-halo mass ratios (see Section~\ref{sec:Galaxygrowth}).

In addition, we compare with the galaxy formation model from UniverseMachine \citep[UM,][]{Behroozi+2019} which has been made available for Shin-Uchuu simulation \citep[][]{Prada+2023}. The resultant mock galaxy catalogue is calibrated to galaxy-halo relationship during forward-modeling to match observational data across cosmic times. The SFRs for these galaxies are calibrated and parametrized as a function of halo mass, halo assembly history, and redshift.

\subsection{BH Growth Model} \label{sec:BHmodel}

The growth of BHs is initiated by the galactic inflows which originate from merger-driven baryonic inflows. The baryonic inflows fuel both the galactic-scale star formation and BH accretion. The growth of BHs has been extensively studied during past decades to account for various BHs from cosmic dawn to local Universe. At cosmic dawn, moderate super-Eddington accretion has been invoked to account for the early existence of supermassive black holes, while in the local Universe, accretion onto supermassive BHs are categorized based on their feedback effect, falling into either ``quasar-mode'' (for Eddington ratio $\lambda>0.01$) or ``radio-mode'' ($\lambda<0.01$) according to their feedback effects \citep[][]{Croton+2006}. Thus, to model the BH growth from cosmic dawn all the way to the local Universe, we adopt BH growth from \citet{Hu+2022b,Hu+2025a} for super-Eddington accretion and adopt phenomenological modeling for sub-Eddington accretion.

At high redshift, to facilitate the rapid growth of young (seed) BHs, super-Eddington accretion is allowed to occur when galactic inflow fueling is sufficient. Following the BH growth model laid out in \citet{Hu+2022b,Hu+2025a}, the BH accretion rate is
\begin{equation}
    \dot{M}_{\rm BH}=\dot{M}_0\cdot \left( \frac{r_{\rm min}}{r_{\rm tr}} \right)^{p},
    \label{eq:BHModel}
\end{equation}
with $r_{\rm tr} \equiv 5\dot{M}_0/(3\dot{M}_{\rm Edd}) \cdot r_{\rm min}$ the photon-trapping radius. We adopt a stellar-mass-dependent index $p$ as indicated by \citep[][]{Hu+2025a}:
\begin{equation}
    p = p_0 - \delta p \cdot \tanh [1.5 \log (M_\star/M_{\rm p})],
    \label{eq:pModel}
\end{equation}
with $M_{\rm p} = 10^9~M_\odot$, $p_0 = 0.75$ and $\delta p=0.25$.
Here, we have implicitly assume that the accretion starts at $r_{\rm min}\equiv 6GM_{\rm BH}/c^2$ for a non-spinning BH. The galactic mass inflow rate into the nuclear region is related to the evolution of host DM halo:
\begin{equation}
    \dot{M}_0 = \mathcal{F} \cdot f_{\rm b} \cdot \frac{{\rm d}M_{\rm h}}{{\rm dt}},
    \label{eq:binflow}
\end{equation}
where $M_{\rm h}$ is halo mass, $\mathcal{F}$ denotes the galactic feeding efficiency, and $f_{\rm b}$ is the baryonic matter fraction. 

\begin{deluxetable*}{rllrll}
\digitalasset
\tablewidth{0pt}
\tablecaption{High-$z$ JWST SMBHs (including some LRDs) discussed in this paper
\label{tab:LRDSample}}
\tablehead{
\colhead{No.} & \colhead{ID} & \colhead{Reference} & \colhead{No.} & \colhead{ID} & \colhead{Reference}
}
\startdata
1 & ZS7 & \citet{Ubler+2024}                    & 14 & J1120+0641   & \citet{Stone+2024}  \\
2 & {UNCOVER-20466} & \citet{Jones+2025}        & 15 & J1030+0524   & \citet{Yue+2024}  \\
3 & \BlueTxt{UHZ1} & \citet{Bogdan+2024}        & 16 & J159-02      & \citet{Yue+2024}  \\
4 & PRIMER-UDS 119639 & \citet{Kocevski+2025}   & 17 & \RedTxt{J0100+2802}   & \citet{Yue+2024}  \\
5 & PRIMER-UDS 29881 & \citet{Kocevski+2025}    & 18 & \BlueTxt{GN-z11}       & \citet{Maiolino+2024b}  \\
6 & UNCOVER-41225 & \citet{Greene+2024}         & 19 & GNZ9         & \citet{Napolitano+2025} \\
7 & UNCOVER-35488 & \citet{Greene+2024}         & 20 & GN-1001830   & \citet{Juodzbalis+2024} \\
8 & UNCOVER-13821 & \citet{Greene+2024}         & 21 & CEERS-1019   & \citet{Larson+2023} \\
9 & JADES-954 & \citet{Maiolino+2024}           & 22 & CEERS-00717  & \citet{Harikane+2023} \\
10 & J2255+0251  & \citet{Ding+2023}            & 23 & \BlueTxt{CEERS-20496}   & \citet{Kocevski+2025} \\
11 & J2239+0207 & \citet{Stone+2024}            & 24 & CEERS-10444   & \citet{Kocevski+2025} \\
12 & J2236+0032 & \citet{Ding+2023}             & 25 & CEERS-7902    & \citet{Kocevski+2025}  \\
13 & J1148+5251 & \citet{Yue+2024}              & 26 & \BlueTxt{Abell2744-QSO1} & \citet{Juodzbalis+2025}
\enddata
\tablecomments{The second and fifth columns list IDs for selected sample, while third and sixth columns list the prime reference for these objects. For instance, for GNz-11, the BH mass is estimated in \citet{Maiolino+2024b}, while the stellar mass of host galaxy is estimated in \citet{Tacchella+2023}.
The four blue-colored objects are selected for showcasing in Section~\ref{sec:LRDs} due to their distinct MM ratios. The rarest (in this case, for DM halos with least merger trees) object J0100+2802 is highlighted in red.
}
\end{deluxetable*}

At lower redshift when accretion goes into low Eddington ratio regime, the BH accretion follows either a ``quasar-mode'' or the ``radio-mode'', as motivated by their feedback dichotomy. In quasar-mode, the accretion follows a log-normal distribution $N(-0.92,~0.46)$ as revealed for low redshift AGN \citep[][]{Schulze+2010}. In ``radio-mode'', the accretion follows a similar distribution but with $10^{-4}$ times lower normalization, i.e., $(-4.92,~0.46)$. 
The active (``quasar-mode'' \& super-Eddington) fraction of SMBHs follows that for low redshift ($1<z<2$) SMBH populations \citep[][]{Schulze+2015} as fitted using
\begin{equation}
      f(M_\star,z) = 10^{k_{\rm BH}\cdot(\log M_\star - b_{\rm BH}) - 2 }
           \left(\frac{1+z}{1+z_{{\rm ref}}}\right)^{\gamma},
\end{equation}
with $k_{\rm BH}=-0.7$ for $\log M_\star<9$ and $k=-2.0$ for $\log M_\star\geq9$, and $ b_{\rm BH}=9.0$, $z_{\rm ref}=1.5$, $\gamma=1.7$. The comparisons between this function form of active fraction and observational indications are shown in Figure~\ref{fig:QFAF}. We caution that a turning point for the active fraction is indicated in observation around $M_{\rm BH}\sim 10^9~M_\odot$. Given the observational uncertainties, this function form fairly capture the broad dependence on redshift and on mass for $z\leq 2$. Although it has been realized that active fraction is higher at higher redshift, no consensus has been reached for $z>3$ so far. To this concern, we simply extrapolate this active fraction to high redshift but capped to $f(M_\star,z)<0.5$. This artifact slow down the rapid growth of seed BHs at high redshift \citep[super-Eddington prescription from][]{Hu+2022b}, which is necessary to explain the existence of $10^9~M_\odot$ SMBHs beyond $z=7$. However, since we focus only on the subsequent evolution for these high-$z$ SMBHs, this artifact does not alter our conclusion significantly.

\subsection{The Modeling Process} \label{sec:process}

In this paper, we pick up $26$ high-redshift ($z\geq 6$) JWST ``overmassive'' SMBHs and LRDs as our initial conditions to investigate their possible evolution down to $z=0$. The whole sample is summarized in Table~\ref{tab:LRDSample}.

\begin{figure}[t!]
    \centering
    \includegraphics[width=0.85\linewidth]{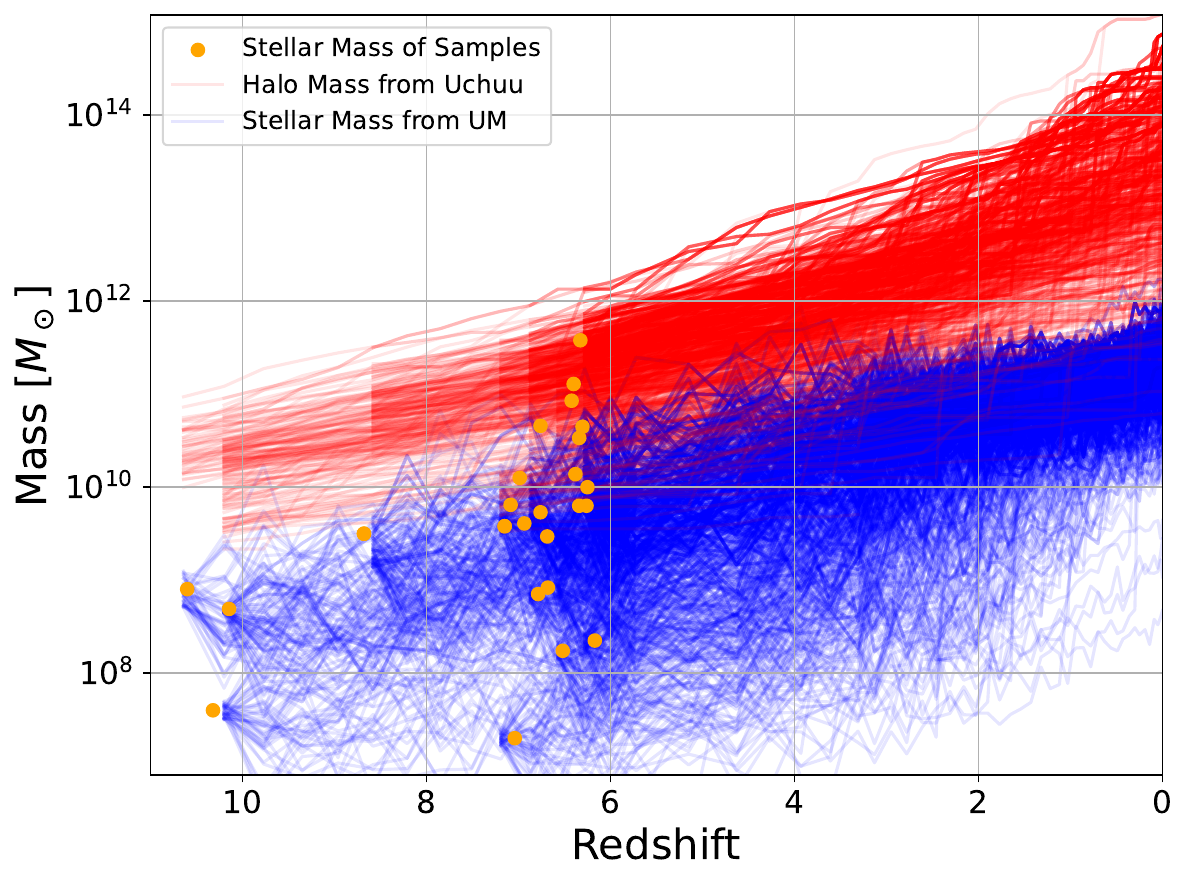}
    \caption{The mass evolution for selected halos and corresponding galaxy model from UniverseMachine, with stellar mass of SMBH and LRD sample overlaid in filled orange circles. DM halo merger trees are selected based on stellar mass criterion with a variation $\pm 10\%$. For each object, $50$ merger trees are selected. For some high-stellar-mass samples with fewer merger trees, the selection criterion is eased to variation of $1$ dex to reach $50$ merger trees. Note for object J0100+2802, only $9$ merger trees are extracted.}
    \label{fig:MergerTrees}
\end{figure}

As a first step, we extract appropriate dark matter halo merger trees from Shin-Uchuu simulation based on Uchuu-UniverseMachine data set \citep[][]{Aung+2023}. The observables for high-$z$ SMBHs serve as the extraction criteria. Since host halo masses are not always available in observation, we adopt stellar masses of host galaxies together with their redshifts as our selection criteria to extract merger trees. During the extraction, the criterion for stellar mass is allowed to vary within $\pm10\%$ of the observed values for all object except rare objects (with less merger trees). For rare objects, the criterion is further eased to allow for variation within $1$ dex. By this approach, we successfully extract $50$ merger trees for most objects except for J0100+2802 \citep[][]{Yue+2024} for which only nine merger tress are identified even for the loose criterion \footnote{The numbers of merger trees are different for different objects. However, for fair comparison and for the sake of clarity, we limit the merger tree number to $50$ for all objects except for J0100+2802 for which the number is nine.}. The extracted merger trees for all selected samples are visualized in Figure~\ref{fig:MergerTrees}.
We note that since Shin-Uchuu simulation data is restored as snapshots for various redshift bins from $z=20$ to $z=0$, initial redshifts for merger trees differ slightly from the precise values from observations. Although
this variation is unavoidable given the limited resolution of redshift bins, it places little effects to our modeling and discussions.

After the extraction, we rebin the halo merger trees to a uniform time resolution ($\delta t_{\rm MT}=10~{\rm Myr}$) to ensure consistent temporal sampling. This time resolution is motivated by the typical life-time of a quasar cycle \citep[][]{Martini+2004}, enabling a reasonable characterization of BH growth. 
In each timestep, we grow the BH and the galaxy simultaneously following the evolution of DM halo merger histories. Despite the accretion and star formation constraints from our phenomenological modeling, their growth rates are limited to the total baryonic inflow rates (the inflow rate $ f_{\rm b} \cdot{{\rm d}M_{\rm h}}/{{\rm dt}}$ in Eq.~\ref{eq:binflow}). This modeling has enabled us to statistically investigate the co-evolution scenario between ``overmassive'' high-$z$ SMBHs and their hosts all the way down to $z=0$ using state-of-the-art N-body simulations.

\section{All Roads Lead to ``Local Relations'': The destinies of overmassive BHs}\label{sec:LRDs}
In the previous section, we developed a semi-analytical approach to model the evolution of both BHs and host galaxies based on DM halo merger trees extracted from N-body simulations. In this section, we apply this model to investigate possible subsequent evolutions of high-$z$ overmassive SMBHs and LRDs and discuss their implications for BH modeling and galaxy formation. We first showcase the evolutionary paths of BHs and host galaxies for four selected high-$z$ SMBHs with different levels of ``overmassiveness''. We demonstrate that despite their various initial MM ratios, the four selected sources exhibit common evolutionary destinations in the MM diagram. Then, we apply the same procedures for all $26$ high-$z$ SMBHs and reveal their similarity in the evolutionary trend except for the rarest object J0100+2802.

\begin{figure*}[t!]
    \centering
    \includegraphics[width=0.95\linewidth]{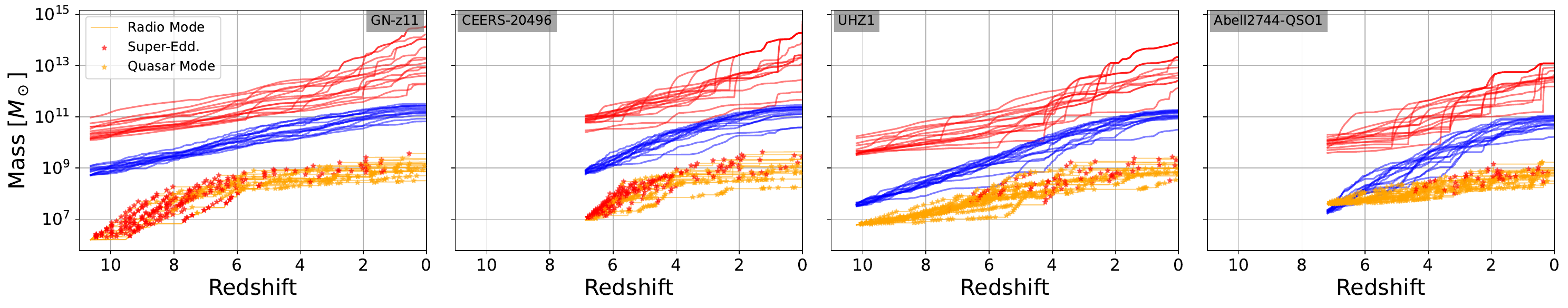}
    \caption{The mass evolution for the selected four sources with different levels of ``Overmassiveness'' (from left to tight): GN-z11, CEERS-20496, UNZ1 and Abell-2744-QSO1. The red, blue and orange colors represent mass evolution for DM halos, for galaxies and for BHs. For BH mass evolution, accretion epochs with super-Eddington rates and at ``Quasar'' mode are highlighted in red and orange stars, while rest epoche are at ``Radio'' mode.
    For better visualization, only $15$ trees out of $50$ are shown in these plots. }
    \label{fig:MassEvolution4}
\end{figure*}

\subsection{The Growth of BHs}\label{sec:BHgrowth}

The BH growth model has been laid out in Section~\ref{sec:BHmodel}, taking into account of BH accretion at various Eddington ratio regimes. Here, we showcase the BH growth for $4$ selected SMBHs with different levels of MM ratios from $\sim 1$ to $\sim 0.1\%$, for GN-z11 \citep[][]{Tacchella+2023,Maiolino+2024b}, CEERS-20496 \citep[][]{Kocevski+2025}, UHZ1 \citep[][]{Bogdan+2024} and Abell2744-QSO1 \citep[][]{Juodzbalis+2025}. 

The mass evolution for BHs and their hosting galaxies and halos are shown in Figure~\ref{fig:MassEvolution4} in orange, blue and red curves respectively for the four objects (from left to right). For BH growth tracks, epochs when accretion rate falls into super-Eddington regime and quasar modes are highlighted in red and orange stars. 
In the beginning, BHs are initiated with similar masses $\sim 10^7~M_\odot$ but seeded inside different halos at different redshifts. These difference lead to diverse evolution of BHs. For instance, the object GN-z11, hosting a relatively small BH compared to its host galaxy, goes through rapid BH growth with frequent super-Eddington accretion during $z=10$ to $z=6$ when BH is small and galactic inflows are sufficient (a necessary condition for the occurrence of super-Eddington accretion, see \citealt{Inayoshi+2016, Hu+2022b}). This rapid evolution of BH mass leads to the increase of MM ratio towards $z\sim 6$, reaching $M_{\rm BH}/M_\star\sim 0.01$. A similar evolutionary trend is observed for CEERS-20496 whose initial mass ratio is $M_{\rm BH}/M_\star\sim 0.01$. In contrast, different early evolutions are observed for the other two objects with higher initial MM ratio. For UHZ1 and Abell2744-QSO1, accretion onto BHs is limited to either quasar mode ($0.01\leq \lambda \leq 1$) or radio mode ($\lambda \leq 0.01$). This early stunted evolution leads to slight decrease in BH-to-galaxy mass ratios, bringing them from ``overmassive'' region (close) to local relations. 

For all four objects, during the late evolution $z\leq 4$, BH evolution enters relative tender phases due to the decrease of active fractions towards lower redshifts. However, we note that even during the tender phases, super-Eddington accretion still occurs occasionally. These super-Eddington accretion episodes occurs preferentially after merger events happen which is believed to facilitate galactic inflows towards nuclear regions in our modeling.
The super-Eddington accreting objects may also be responsible for high luminosity quasars observed at these epochs.

In a word, our BH growth model enables super-Eddington accretion for undermassive BHs, while for overmassive/normal SMBHs, their growth is limited to either quasar mode or radio mode, both bringing BHs closer to local BH-to-galaxy mass ratios. At lower redshift, BH growth tends to be more moderate but super-Eddington accretion is possible and occurs occasionally.

\subsection{The Growth of Host Galaxies}\label{sec:Galaxygrowth}

 For the selected four sources, we showcase their stellar mass evolution in blue curves in Figure~\ref{fig:MassEvolution4}. Due to higher SFR for star-forming main sequence galaxies at higher redshift and relatively lower quiescent fraction, galaxies grow their mass substantially at redshift beyond $z\sim 2$, while the growth is weakened at lower redshifts when they spend most of their time in quiescent stages.

\begin{figure}[t!]
    \centering
    \includegraphics[width=0.8\linewidth]{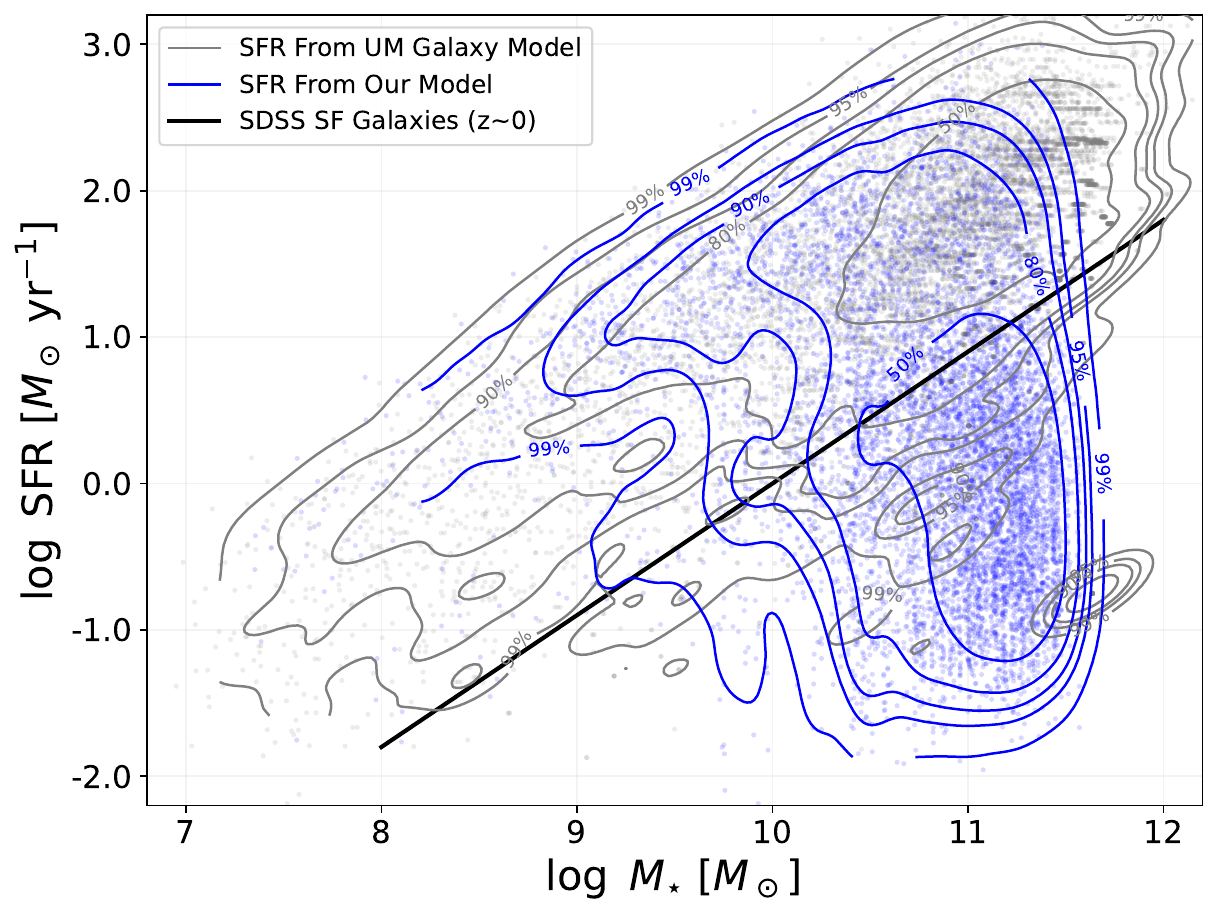}
    \includegraphics[width=0.85\linewidth]{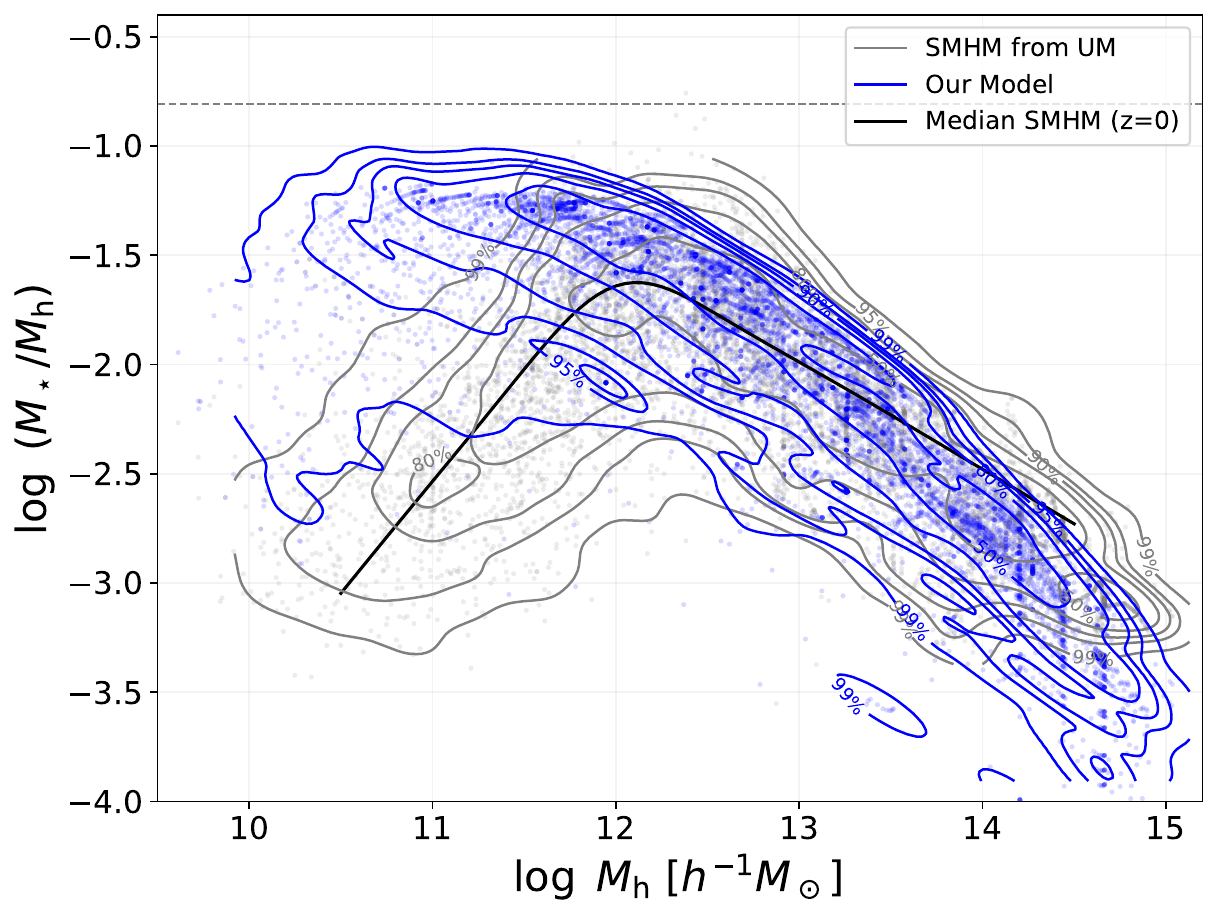}
    \caption{Upper panel: The SFR distribution as a function of stellar mass for our galaxy model (blue dots and blue contours) and for UM galaxy model (gray dots and gray contours). For clarity, only galaxy models for $4$ selected samples are shown in both plots. The black curve shows the fitted relation for $z\sim 0$ galaxy samples from SDSS in \citet{Peng+2010}. 
    Lower panel: The stellar-to-halo-mass (SMHM) ratio for $4$ selected samples, for our model (blue curves) and for galaxy formation model from UniverseMachine (gray curves). The black curve represents the inferred median stellar-to-halo-mass ratio at $z=0$ in UM \citep[][]{Behroozi+2019}.}
    \label{fig:SFRConstraints}
\end{figure}

Meanwhile, we also compare our galaxy formation model with the comprehensive UM galaxy formation model for Shin-Uchuu simulations \citep{Aung+2023}. In Figure~\ref{fig:SFRConstraints}, we present comparisons of galaxy properties between ours (blue colors) and UM galaxy (gray colors) models in terms of SFR distribution (upper panel) and stellar-to-halo-mass ratio (SMHM, lower panel). 

\begin{figure*}[t!]
    \centering
    \includegraphics[width=0.95\linewidth]{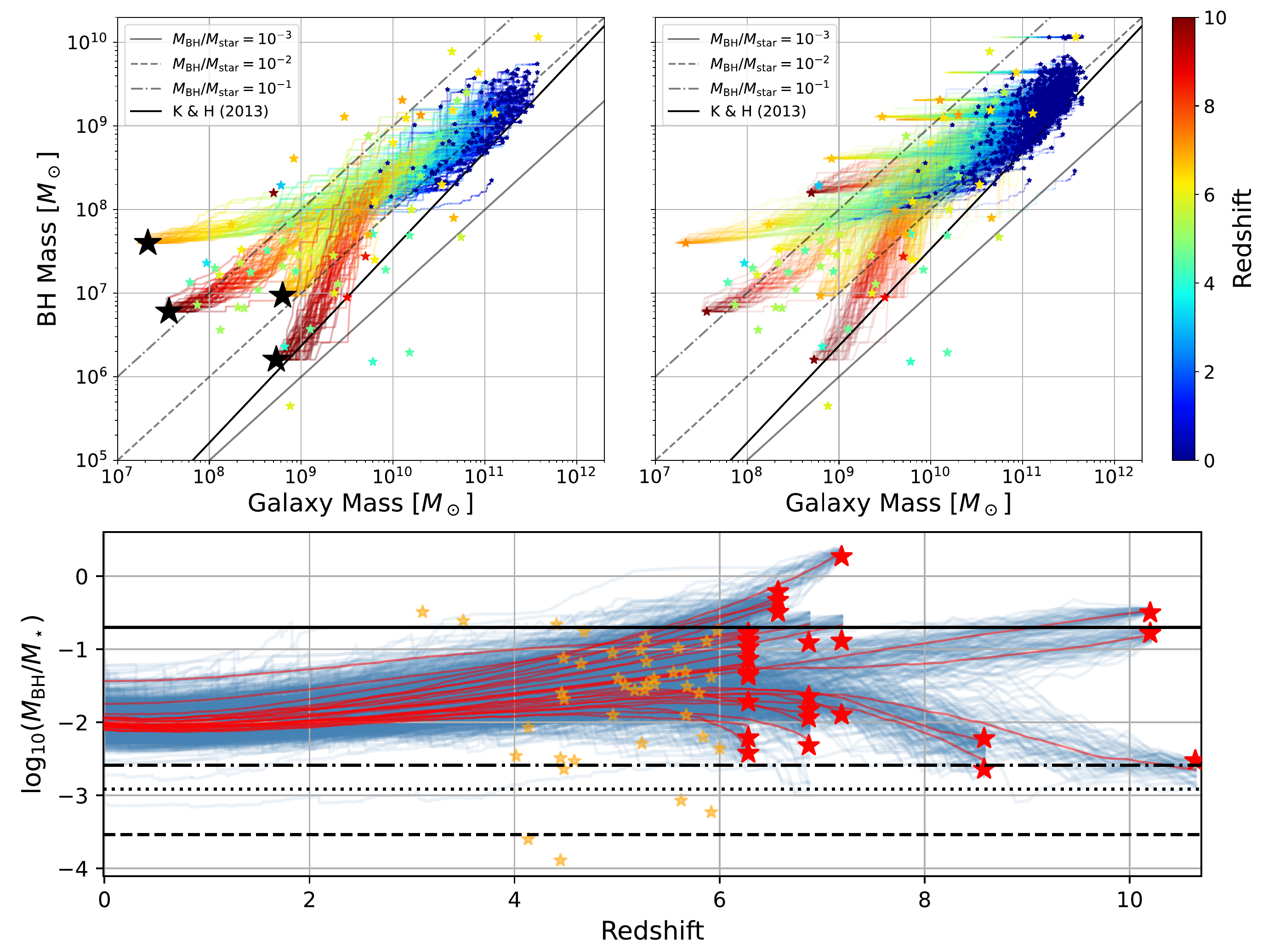}
    \caption{Top panels: The evolution of BH-to-galaxy-mass ratios for the four selected sources (left panel) and for the whole sample (right panel). In top-left panel, initial conditions for the four sources are highlighted in black stars, while their evolutions are shown in solid curves with redshift color-coded. At $z=0$, their evolution destinations are highlighted in blue stars.
    As comparisons, other high-$z$ JWST SMBHs and LRDs \citep[][]{Ding+2023,Harikane+2023,Kocevski+2023,Larson+2023,Ubler+2023,Maiolino+2024,Maiolino+2024b,Juodzbalis+2024,Bogdan+2024,Ubler+2024,Stone+2024,Yue+2024, Greene+2024,Kocevski+2025} are shown in color-coded stars and local relation from \citet{Kormendy+2013} and characteristic ratios ($M_{\rm BH}/M_\star=10^{-3},~10^{-2},~10^{-1}$) are overlain in black curve and in gray curves.
    Lower panel: The MM ratio evolution as a function of redshifts for the whole sample and for other high-$z$ SMBHs and LRDs. The initial conditions for $26$ selected sources are highlighted in red stars. The solid blue curves are individual evolution of MM ratios for all sources, while the solid red curves represent the averaged MM ratio evolution. 
    The dashed, dashed–dotted, and dotted horizontal lines are BH-to-galaxy-mass ratios for late-type, early-type, and all galaxies at  $M_\star=3\times 10^{10}~M_\odot$ \citep[][]{Greene+2020}, respectively. The black solid curve shows the upper limits for MM ratio in BH growth model from \citet{Hu+2025a}, while orange stars are for high-$z$ JWST SMBHs and LRDs.
    }
    \label{fig:MMratio}
\end{figure*}

In the upper panel, SFR distributions as a function of stellar mass are shown for the two modelings with the relation for local SDSS star-forming galaxies overlain in black curve \citep[][]{Peng+2010}. Since in both models galaxies are initiated at high-$z$, the distribution compiles SFR at all redshift epochs other than $z\sim 0$. It has been pointed out that the main sequence SFR relation at high redshift would be higher \citep[e.g., ][]{Speagle+2014,Leja+2022}, which naturally accounts for the disparity between the modeling and the local observation. Despite the differences, a noticeable fraction of our samples follows the local relation, while a large fraction of the samples are in quiescent phase (normally $\sim 1-1.5$ dex lower than the main sequence relation). The transition to the quiescent phase seems smooth, failing to capture the green-valley structure. This might be due to the adopted unphysical bimodal distribution of SFR. 

In the lower panel, we present the comparison of SMHM ratio (equivalent to star formation efficiency when multiples baryonic fraction $f_{\rm b}^{-1}$) for galaxies of the four selected sources. The SMHM relation for UM galaxy formation model has been calibrated to observations across a wide range of redshifts, captures the overall structures of actual SMHM relation: an increase of star formation efficiency at lower mass end and a decrease towards high mass end, with a peak around halo mass $M_{\rm h}\sim 10^{11}-10^{12}~M_\odot$. In comparison, our model captures the high-mass tail evolution well while overshot the lower-mass end relations. In the lower mass end (equivalently at high-$z$ universe), our model predict about one dex higher galaxy mass. One possible reason is that our galaxy samples are hosts for rare objects at high-$z$ where higher stellar formation efficiency has been indicated \citep[e.g.,][]{Harikane+2023,Haro+2023,Wang+2024}.
Nevertheless, these comparisons indicate that our empirical galaxy model captures the major characteristics of galaxy formation, in broad consistency with observations.

\subsection{The Co-evolution of SMBHs and Host Galaxies}\label{sec:BHGalaxyCoevolution}

After the construction of BH and galaxy models, we turn to the investigation of their evolution in MM ratios. In the top-left panel of Figure~\ref{fig:MMratio}, we present the mass ratio evolution for the four selected sources. The initial conditions for the four sources are shown in black stars in the figure. Their subsequent evolution in MM diagram are shown in solid curves with redshift color-coded. At $z=0$, their destinations in the diagram is highlighted in blue stars. As comparisons, local relation from \citet{Kormendy+2013} and characteristic ratios ($M_{\rm BH}/M_\star=10^{-3},~10^{-2},~10^{-1}$) are overlain in black curve and in gray curves, while other high-$z$ JWST SMBHs and LRDs are shown in redshift color-coded stars \citep[][]{Ding+2023,Harikane+2023,Kocevski+2023,Larson+2023,Ubler+2023,Maiolino+2024,Maiolino+2024b,Juodzbalis+2024,Bogdan+2024,Ubler+2024,Stone+2024,Yue+2024, Greene+2024,Kocevski+2025}.

The four samples are selected due to their distinct MM ratios, located at distinct regions in the figure. After ``seeding'', BHs and galaxies grow substantially at different paces. For the overly overmassive two objects, the galaxy growth rate surpasses BH growth rate, gradually lowering the MM ratios. For CEER-20496 ($M_{\rm BH}/M_\star\sim 0.01$), galaxies and BHs grow at similar rates maintaining an almost constant evolution of mass ratio. For GN-z11 ($M_{\rm BH}/M_\star\sim 0.005$), initial super-Eddington accretion episodes bring the mass ratio to $M_{\rm BH}/M_\star\simeq 0.01$ in a short duration, which is roughly maintained to $z=0$. We notice that despite their distinct initial conditions, all evolutionary tracks seem to converge at galaxy mass scale $M_\star\sim 10^{9}-10^{10}~M_\odot$. This convergence behavior is attributed to the ``feedback model'' adopted in Equation~(\ref{eq:pModel}) as noted in \citep[][]{Hu+2025a}. At $z=0$, almost all BHs evolve into local relations as indicated by the blue stars, with a small fraction of BHs being outliers. 

Similarly, we conduct the same analysis for the rest $22$ high-$z$ SMBHs and LRDs. We present their evolutionary tracks in the top-right panel of Figure~\ref{fig:MMratio}. We find that the previously selected four sources bracket almost all possible distributions of initial mass ratios. The conclusions for the four sources still hold for the whole sample. In addition, we notice that for some high-mass-ratio ($M_{\rm BH}/M_\star\geq 0.1$) and extreme mass ($M_{\rm BH}\geq 10^9~M_\odot$) SMBHs,  their evolutionary tracks are flat for the majority of their life times indicative of stunted growth of SMBHs. They likely remain slightly overmassive even at $z=0$. For instance, for the rarest object J0100+2802, no BH growth is observed while still being overmassive at $z=0$ (about one dex higher the local relation).

In the lower panel of Figure~\ref{fig:MMratio}, we show the redshift evolution of MM ratio for all $26$ sources, with their initial conditions overlaid in red stars and other high-$z$ SMBHs and LRDs in orange stars. Despite their origins, the evolution tracks for MM ratios of the whole sample go through ``overmassive'' phases at moderate redshift ($4\lesssim z \lesssim 6$), an epoch coinciding with the peak distribution of LRD samples \citep[e.g.,][]{Inayoshi+2025}, and then converge to local relations with variation about $0.5$ dex. Although the convergence behavior is due to the formulism of ``feedback model'' in BH modeling, the ``overmassive'' evolutionary stages might be inevitable for BHs growing at super-Eddington accretion rates, even episodically. However, if the bulk evolution of these BHs and galaxies evolve according to the statistical properties of BH accretion rates, active fractions, SFR and quiescent fractions (as modeled in this paper), they will finally be back to normal (local relation) again at $z=0$.

\section{Discussions}
\label{sec:Discussion}

Based on observables, we model possible evolutionary tracks for $26$ high-$z$ ($\geq 6$) JWST ``overmassive'' SMBHs and LRDs combining N-body simulations and analytical modeling. Our phenomenological modeling successfully captures the ``overmassive'' evolutionary stages of these SMBHs and LRDs and their subsequent evolution down to $z=0$, approaching local relations. Despite its straightforward formulism, the results of the model turn out to be interesting and indicative.

\subsection{Implications for BH and Galaxy Modeling}
\label{sec:DiscussionBHGalaxy}

Our modeling allows for both super-Eddington and sub-Eddington growth for BH evolution. The super-Eddington accretion is argued to be essential to grow SMBHs at $z\geq 6$ with masses $M_{\rm BH}\geq 10^9~M_\odot$ \citep[e.g.][and references therein]{Inayoshi+2020}, but it also has been reported that some low-$z$ quasars host super-Eddington accreting SMBHs \citep[][]{Collin+2004,Du+2014}. It motivates us to properly include the super-Eddington prescription in BH modeling. As stated in Section~\ref{sec:BHgrowth}, super-Eddington accretion almost dominates the early stage growth for these ``low'' mass ratio BHs where super-Eddington conditions are easily met. This rapid growth outpaces the evolution of their host galaxies and leads to an increase of mass ratio to $M_{\rm BH}/M_\star\sim 0.01$, even to higher mass ratios \citep[][]{Hu+2022b,Hu+2025a}. Combining the ancient and subsequent evolutions for high-$z$ SMBHs, it points to the indication that super-Eddington accretion preferentially occurs for ``low'' mass ratio (seed) BHs and is accountable for the emergence of overmassive SMBHs.

\begin{figure*}[t!]
    \centering
    \includegraphics[width=0.31\linewidth]{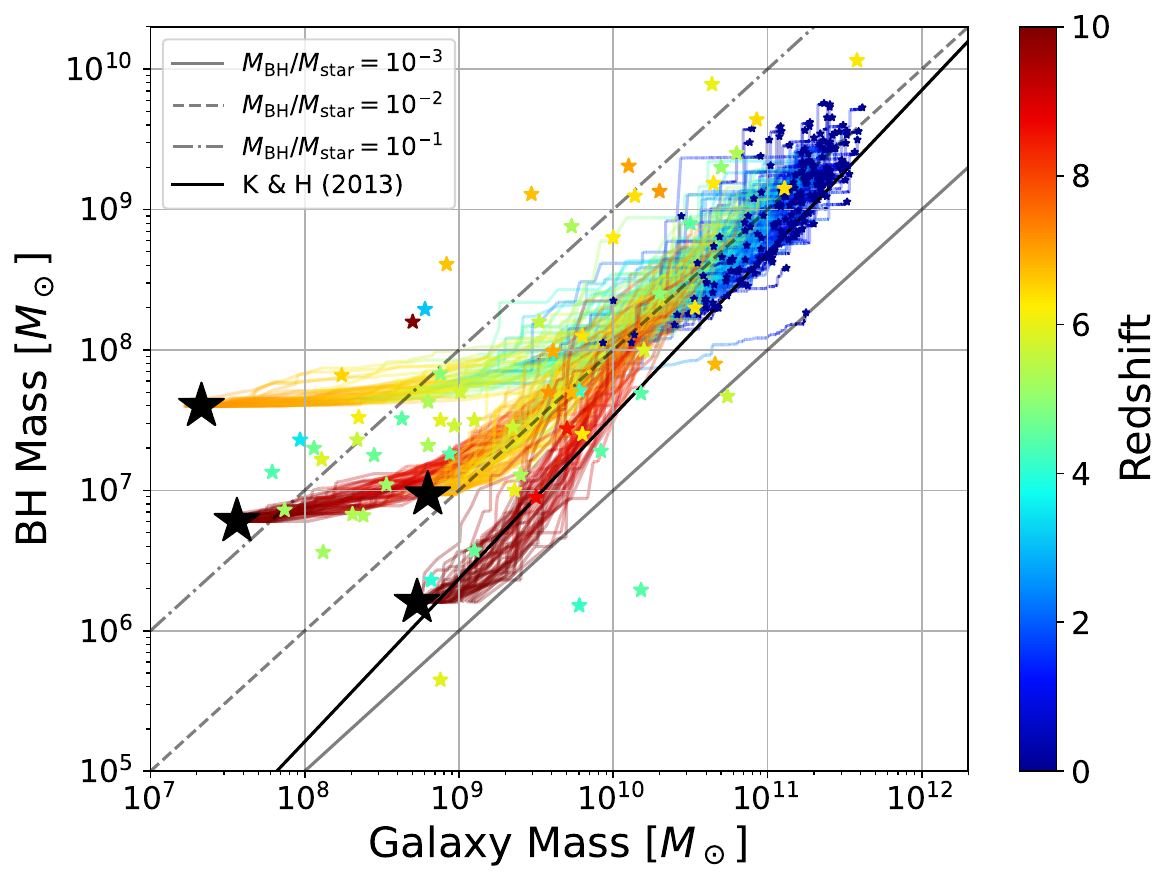}
    \includegraphics[width=0.31\linewidth]{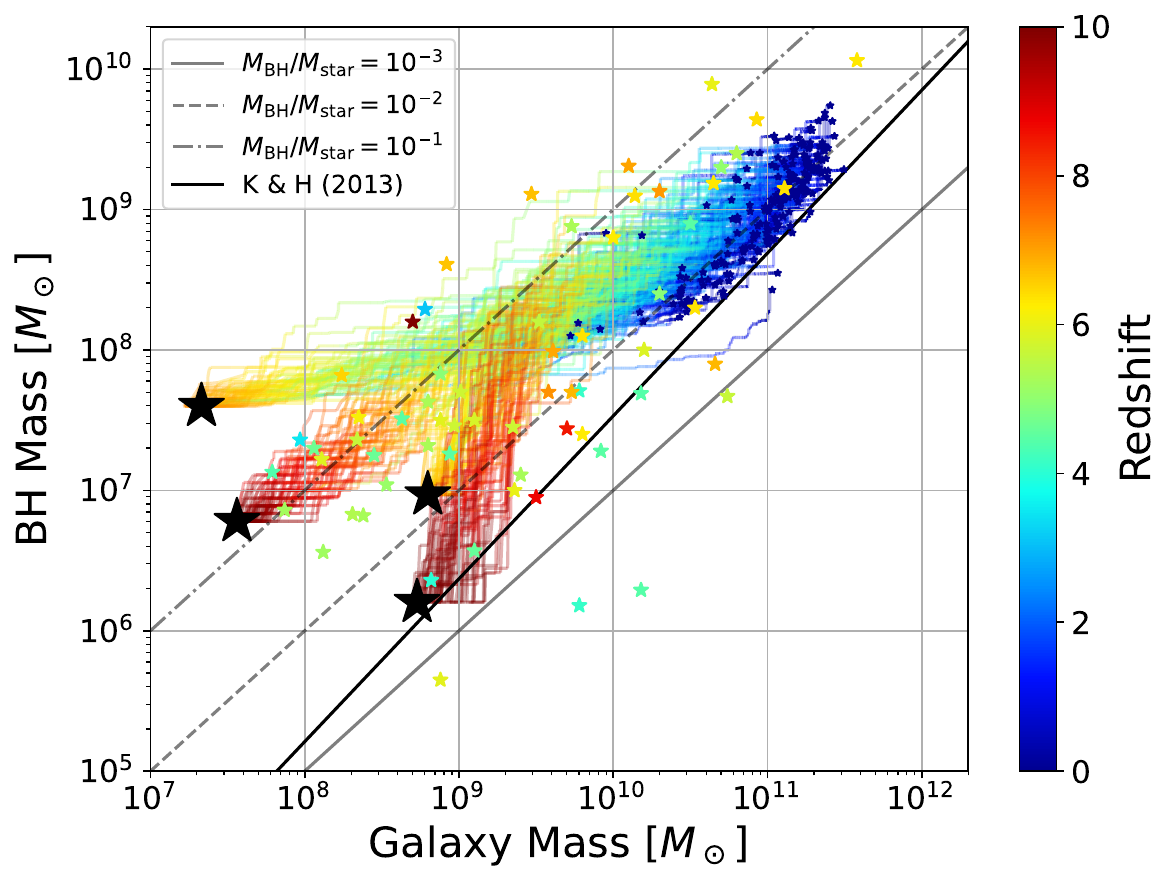}
    \includegraphics[width=0.31\linewidth]{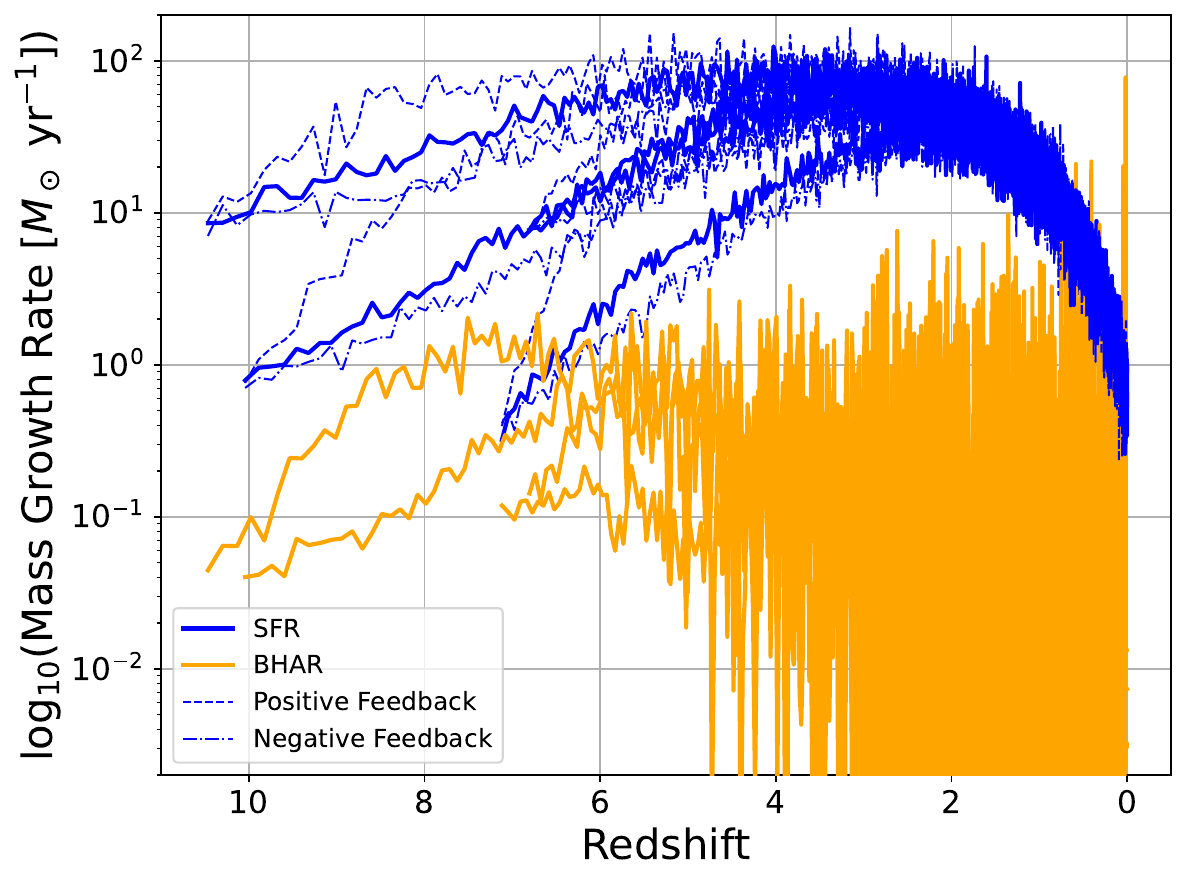}
    \caption{Left \& middle panels: The evolution of BH-to-galaxy-mass ratios of the four selected sources for ``positive'' (left panel) and ``Negative'' (middle panel) feedback models. The symbols for stars and curves are the same as these in Figure~\ref{fig:MMratio}.
    Right panel: The averaged star formation rate (SFR, blue curves) and BH accretion rate (BHAR, orange curves) for fiducial case (solid curves), cases with positive (dashed curved) and negative (dash-dotted curves) feedback. The growth rates are averaged over $50$ merger trees. 
    For BH growth, only fiducial case is shown for clarity as BH growth is statistically the same for all models.
    }
    \label{fig:MMratio2}
\end{figure*}

In the sub-Eddington regime, the BH accretion follows an observation-based dichotomy of ``quasar-mode'' and ``radio-mode'', with an empirical active fraction. This phenomenological modeling is useful and statistically accurate enough to grow SMBHs down to $z\sim 0$, neither starving or overfeeding them. The low-accretion evolution dominates the low redshift ($0\lesssim z \lesssim 4$) evolution and finally brings mass ratios of (majority) high-$z$ SMBHs and LRDs to local relations. On the one hand, it indicates that the observational relations of Eddington ratio distribution and active fraction of SMBHs are valid to a large extent. Therefore, it requires reasonable explanations for these empirical relations during simulations and theoretical modelings, further appealing for a better understanding of the mutual feedback mechanism as a whole. In addition, sub-Eddington accretion occurs only when nuclear-scale galactic inflows, $\dot{M}_0$ in Equation~(\ref{eq:BHModel}), are insufficient to sustain super-Eddington accretion. This process is tightly related to how galactic scale gas is fed to nuclear BHs, which remains poorly constrained so far.

As for the galaxy evolution, our empirical modeling captures the mass evolution trend and major characteristics, e.g., SFR distribution and star-formation efficiency distribution, despite all details are neglected. Similar to BH modeling, these simplified relations are useful and awaiting a more straightforward theoretical explanation. Meanwhile, we caution that though ``feedback'' is not implemented, its mutual effects are somehow coupled to the distribution of growth rates (accretion \& SFR) and active (quiescent) fractions. Nevertheless, we experiment with ``positive'' and ``negative'' feedback modeling to investigate their possible signatures in the evolutionary tracks.

\subsection{Positive Feedback vs Negative Feedback}
\label{sec:PositiveFeedback}

One of the advantages of our modeling is the freedom to connect the BH accretion and star-formation in different manners. To shed lights on feedback mechanism, we conduct experiments to positively and negatively correlate star-formation to BH accretion, namely ``positive feedback'' and ``negative feedback''. In the canonical dichotomy feedback model, i.e., BH accretion feedback (aka, AGN feedback) acts either to positively enhance or to negatively suppress galactic star-forming. In practice, we manually increase/decrease the SFR by a factor of $f_{\rm fd}=5$ for each time epochs when BH accretion occurs at high rates, i.e., super-Eddington accretion and ``quasar mode'' (Eddington ratio $\lambda \geq 0.01$). Although the feedback strength varies when BH accretion rate changes, the factor $f_{\rm fd}=5$ is found to be large enough to manifest its effects on the evolution of MM ratios. No time-delay between star-formation and BH accretion is considered, since statistical average is always expected in our modeling. 

In the left and middle panels of Figure~\ref{fig:MMratio2}, we present the MM ratio evolutionary tracks for ``positive feedback'' (left) and ``negative feedback'' (middle). For a clear visualization, we only show the mass ratios for the selected $4$ samples as did in Section~\ref{sec:LRDs}. These two plots are exactly the same as the top-left panel in Figure~\ref{fig:MMratio}, except for different feedback models. Comparing with Figure~\ref{fig:MMratio}, we find there are significant differences between different feedback models during their early evolution, though they are found to converge to local relations at $z=0$ despite feedback models.

For ``positive feedback'', super-Eddington accretion at early stages also boosts the galaxy formation for ``low ratio'' objects (for GN-z11 and CEERS-20496 in this case), leading to a regulated evolution of mass ratios all the way down to local relation at $z=0$. For the two ``overmassive'' objects, galaxy formation is accelerated moderately while BH growth is maintained as usual, resulting in rapid convergence of mass ratios to local relations. For ``negative feedback'', super-Eddington accretion acts to suppress the star-formation for ``low ratio'' objects, leading to a rapid increase of mass ratio at early stages. On the other hand, the two ``overmassive'' objects whose early stages are dominated by sub-Eddington accretions, experience significant growth of stellar mass, maintaining (or slightly decrease) the mass ratio around (to) $0.01-0.1$. We also notice that despite feedback models, all BHs share a common evolutionary end, converging to the local relations. The common evolutionary end seems due to the phenomenological modeling which adopts observational-based empirical relations for BH accretion and SFR. It indicates that all feedback signatures have already been washed out by statistical averages. To disentangle this, we could either look for any redshift evolution for the mass ratios at higher redshifts or decipher the feedback effects from the empirical relations. We expect efforts from both observations and theoretical modelings.

For quantitative comparison, the averaged growth rates for BHs and for host galaxies are shown in the right panel of Figure~\ref{fig:MMratio2}. In this panel, the averaged BH growth rates are shown in orange curves for the $4$ selected samples, while the averaged host galaxy growth rates are shown in blue curves for fiducial (solid), positive (dashed) and negative (dash-dotted) feedback models respectively. For BH growth, only fiducial case is shown for clarity as BH growth is statistically the same for all models. The feedback effects have been manifested in the evolution of galaxy growth rates. As expected, the galaxy growth is boosted for the positive feedback cases while being suppressed for the negative feedback case during their early evolution ($z>4$). However, the boosting and suppressing fractions are different for the two cases. For positive feedback model, the averaged growth rate for host galaxies can be $10$ times as large as the fiducial case, while for negative feedback model, the growth rate is only suppressed by a factor of $\sim 2$ (smaller than the input suppression factor $f_{\rm fd}=5$). The difference is largely attributed to the statistical effect. It indicates that positive effects (if exist) may be boosted while negative effects may be suppressed in statistical analysis.

\section{Conclusions}
\label{sec:conclusions}

In this paper, we investigate the possible subsequent evolutionary tracks for high redshift ($z>6$) ``overmassive'' supermassive black holes and ``Little Red Dots'' based on dark matter halo merger histories from N-body simulations and empirical relations for BH accretion and star formation, allowing for super-Eddington accretion. The dark matter halo mergers provide the fundamental laboratories and bring in baryonic gas for the formation and evolution of BHs and galaxies. In our modeling, the galactic gas inflows fuel the star-formation in an empirical manner and a fraction of inflows is fed to nuclear BHs. When inflows are sufficient, BHs are fed at super-Eddington rates, otherwise at sub-Eddington rates.
For both galaxy and BH growth at sub-Eddington regime, observational based empirical relations are utilized as constraints for their evolution at different mass scales and at different redshifts. We successfully model the subsequent evolution for $26$ high-$z$ SMBHs and LRDs to $z=0$. Our main findings are summarized as follows:

\begin{enumerate}
\item For ``low'' mass ratio SMBHs and LRDs, super-Eddington accretion dominates their early stage evolution, which outpaces the evolution of their host galaxies and leads to increase of their mass ratio to $M_{\rm BH}/M_\star\sim 0.01$. Afterwards, sub-Eddington accretion takes over and regulates BH evolution to local relations at $z=0$.
\item For most of overmassive SMBHs and LRDs, sub-Eddington accretion always dominates the evolution of BHs. In this case, BH growth is not completely stunted and accretion gradually brings these BHs back to local relations. 
\item In almost all cases, the mass ratio goes through an ``overmassive'' phase either due to initial conditions or early rapid evolution for BHs. BHs at these phases naturally explains the existences of the ``overmassive'' SMBHs and LRDs. Additionally, super-Eddington accretion occurs occasionally when conditions are met. In these occasions, highly accreting SMBHs might reveal their existences as luminous quasars. 
\item For some rare objects, they may never go back to local relations again. The reasons are twofold. They are mature enough that almost no significant BH growth is observed. Meanwhile, they are too rare that only limited merger events occur during their halo evolution, resulting in limited fueling available for BH growth and star-formation.  

\item Our feedback experiments point out that the feedback effects are washed out at $z=0$ due to our phenomenological modeling. Nevertheless, our modeling guide us to decipher the feedback effects from both the possible redshift evolution of mass ratios at higher redshifts and local empirical relations.
\end{enumerate}

Finally, we caution that despite our successful modeling, some prescriptions are simplified and empirical-based, e.g., the galactic feeding efficiency, mass loss due to stellar evolution. It brings in the major uncertainties, complicating our interpretations. Though the empirical relations are powerful and straightforward for our modeling, it impedes a deeper interpretations for the co-evolution between SMBHs and host galaxies. Physically motivated investigation for these relations are awaiting for further study.

\begin{acknowledgments}
We thank the organizers of the Galaxy-IGM Workshop 2025 for hosting an inspiring meeting where the primary idea for this paper was developed and thank Suzuka Arai for discussions. 
This research was supported by the Japan Society for the Promotion of Science (JSPS) through KAKENHI Grant Numbers 24KF0130 (HH, KO), JP21H04488(KO),  25K01045 (KO). This work was also supported by MEXT as ``Program for Promoting Researches on the Supercomputer Fugaku'' (Structure and Evolution of the Universe Unraveled by Fusion of Simulation and AI; Grant Number JPMXP1020240219; KO), by Joint Institute for Computational Fundamental Science (JICFuS, KO), and (in part) by the Multidisciplinary Cooperative Research Program in CCS, University of Tsukuba, and supported by IAAR Research Support Program in Chiba University Japan (TI). 
HY acknowledges support by KAKENHI (25KJ0832) through Japan Society for the Promotion of Science (JSPS).
TI acknowledges support from JPMXP1020240219 and JICFus. The numerical calculations were performed partially on HPE Cray XD2000 at the Center for Computational Astrophysics (CfCA), National Astronomical Observatory of Japan and partially on Yukawa-21 at Yukawa Institute for Theoretical Physics (YITP) in Kyoto University. 

We also thank Instituto de Astrofisica de Andalucia (IAA-CSIC), Centro de Supercomputacion de Galicia (CESGA) and the Spanish academic and research network (RedIRIS) in Spain for hosting Uchuu DR1, DR2 and DR3 in the Skies \& Universes site for cosmological simulations. The Uchuu simulations were carried out on Aterui II supercomputer at Center for Computational Astrophysics, CfCA, of National Astronomical Observatory of Japan, and the K computer at the RIKEN Advanced Institute for Computational Science. The Uchuu Data Releases efforts have made use of the skun@IAA\_RedIRIS and skun6@IAA computer facilities managed by the IAA-CSIC in Spain (MICINN EU-Feder grant EQC2018-004366-P).
\end{acknowledgments}

\bibliography{sample701}{}
\bibliographystyle{aasjournalv7}
\end{CJK*}
\end{document}